\documentclass[a4,11pt]{article}
\usepackage{amssymb}
\usepackage{amsmath}
\usepackage[pdftex]{graphicx}
\usepackage{cite}
\usepackage{slashed}
\usepackage{bm}
\usepackage[top=2.5cm, bottom=2.0cm, left=2.5cm, right=2cm]{geometry}

\begin{document}

\begin{center}
{\large {\bf The contribution of new physics on the exclusive W boson hadronic decays in the final state at muon colliders in the Randall-Sundrum model }}
\\

\vspace*{1cm}

{ Bui Thi Ha Giang$^{a, }$ \footnote{e-mail: giangbth@hnue.edu.vn}, Dang Van Soa$^{b }$}\\

\vspace*{0.5cm}
 $^a$ Hanoi National University of Education, 136 Xuan Thuy, Hanoi, Vietnam\\
 $^b$ Faculty of Applied Sciences, University of Economics - Technology for Industries,\\
 456 Minh Khai, Hai Ba Trung, Hanoi, Vietnam
\end{center}

\begin{abstract}
	
\hspace*{1cm}
An attempt is made to present the effect of new physics in the exclusive decays of W boson at high energy colliders in the Randall-Sundrum (RS) model. By using Feynman diagram techniques we have evaluated the influence of the scalar unparticle, polarization and anomalous couplings on the exclusive W boson hadronic decays of $W^{\pm} \rightarrow \pi^{\pm}\gamma$, $W^{\pm} \rightarrow K^{\pm}\gamma$ and  $W^{\pm} \rightarrow \rho^{\pm}\gamma$ at the high energy muon colliders in the RS model. The result shows that with fixed collision energies, the total cross-section for hadronic productions in the final state depends strongly on the parameters of the unparticle physics, muon beam polarizes and also  anomalous couplings. With a center-of-mass energy of 10 TeV, the total cross-sections achieve the maximum value when the benchmark signal point as $(\Lambda_{U}, d_{U})$ $= (1 \text{TeV}, 1.9)$ and the polarization coefficient as $(P_{\mu^{-}}, P_{\mu^{+}} )= (1,1)$. The numerical evaluation shows that with the contribution of new physics in the RS model, the effect is greatly enhanced at high energy colliders. The statistical significance can reach $1\sigma - 7\sigma$ in case of using the W branching ratio at the experimental bounds of $\mathcal{O} (10^{-6})$.  To clarify the contribution of new physics, we use a $\chi^{2}$ analysis with systematic errors to determine the sensitivities of the new contributions. The result indicates  that the sensitivities on the anomalous coupling $WW\gamma$ are much larger than that on the anomalous coupling $WW Z$ under the same conditions.\\
\end{abstract}


\section{Introduction}
\hspace*{1cm}One of the most attractive extended models beyond the Standard model (SM) is the Randall-Sundrum (RS) model to solve the hierarchy problem  naturally \cite{rs}. The RS model setup involves two 3-branes bounding a slice of 5D compact anti-de Sitter  space. Gravity is localized UV brane, while the SM fields are supposed to be localized IR brane. The separation between the two 3-branes leads directly to the existence of an additional scalar called the radion ($\phi$), corresponding to the quantum fluctuations of the distance between the two 3-branes \cite{csa, gold, dominici}. Radion and Higgs boson have the same quantum numbers, the general covariance allows a possibility of mixing between the radion and the Higgs boson \cite{dominici, ahm}. Phenomenology of scalar particles (Higgs/radion) at the high energy colliders in the RS model has been intensively studied in Refs.\cite{dav, soa, soa1, ali, bha, boo, soael}.\\
\hspace*{1cm} At TeV scale, the scale invariant sector has been considered as an effective theory and that if it exists, it is made of unparticle suggested by Geogri \cite{georgi,georgi2} and may become part of reality. The very high energy theory contains the fields of the SM and the fields of a theory with a nontrivial IR fixed point which is called Banks-Zaks (BZ) fields \cite{banks}. These have the generic form between the SM and BZ fields as follows \cite{tmaliev, chenhe}. 
\begin{equation}\label{ptu1}
\dfrac{1}{M_{U}^{k}}\mathcal{O}_{SM}\mathcal{O}_{BZ},
\end{equation} 
where $\mathcal{O}_{SM}$ is an operator with mass dimension $d_{SM}$ built out of SM fields and $\mathcal{O}_{BZ}$ is an operator with mass dimension $d_{BZ}$ built out of BZ fields. In the effective theory below the scale $\Lambda_{U}$, the BZ operators match onto unparticle operators and the interactions of (\ref{ptu1}) match onto interactions of the form
\begin{equation}\label{ptu2}
\dfrac{C_{U}\Lambda_{U}^{d_{BZ} - d_{U}}}{M_{U}^{k}}\mathcal{O}_{SM}\mathcal{O}_{U},
\end{equation}
here $d_{U}$ is the scaling dimension of the unparticle operator $\mathcal{O}_{U}$. Unitarity constraints derived from Conformal Field Theory (CFT) set the fundamental lower bound on the dimension at $d_{U} \geq 1$. However, operators with dimensions $d_{U} \leq 2$ are preferentially considered, exceeding this threshold ($d_{U} \geq 2$) makes the calculations unpredictable due to an increased sensitivity to the ultraviolet sector. Therefore, the range $1< d_{U} <2$ is considered \cite{frass}. Based on the BZ theory \cite{banks,chenhe}, unparticle stuff with nontrivial scaling dimension is considered to exist in our world and this opens a window to test the effects of the possible scalar invariant sector, experimentally. The effects of unparticle on properties of high energy colliders have been intensively studied in Refs.\cite{cheung, pra, alan, maj, kuma, sahi, kiku, chen, kha, fried, iltan, giang}. Dark matter and unparticles production in association with a Z boson in pp collisions at the center-of-mass energy 8 TeV has been considered in Ref.\cite{khac}. Recently, the scalar unparticle signals at LHC are studied in detail in Ref.\cite{alie}. The anomalous couplings at LHeC is researched in Ref.\cite{giang2023}.\\
\hspace*{1cm} The muon colliders, which can reach center-of-mass energy up to tens of TeV, provide an unprecedented potential in probing new physics beyond the SM\cite{blas}. The advantage of initial muon beam polarization is that it is effective for the indirect search \cite{fuku, ska}, which is versatile in the fundamental particle physics and nuclear physics \cite{cohen}. The integrated luminosity scaling of a high energy muon collider (assuming a 5-year run) reaches 1 $\text{ab}^{-1}$ (3 TeV), 10 $\text{ab}^{-1}$ (10 TeV) and 20 $\text{ab}^{-1}$ (14 TeV)\cite{asadi, cap, wulzer, liu}. Given that the radion mixes with the Higgs boson, precision measurements of Higgs properties at muon colliders will provide crucial information \cite{li, fors}. In our previous work, investigation of the scalar unparticle and anomalous couplings at muon colliders in final states with multiple photons in the RS model is considered in detail in Ref.\cite{gs2024}. The contribution of the scalar unparticle on the WW production at ILC is studied in Ref.\cite{giangMPLA} in which we only considered in s-channel without considering the influence of $\gamma WW, ZWW$ anomalous couplings and neutrino propagator. Moreover, the rare hadronic decays of W bosons in the final states would provide an accurate measurement of the W boson mass that is based solely on visible decay products at future colliders \cite{siru1, siru2, gaad}. These rare processes, if observed, would validate the quantum chromodynamics (QCD) factorization formalism used to calculate cross-sections at colliders \cite{gaad}. In our work, we consider the  exclusive decays of W boson into a photon plus a charged meson. It is worth that there is not universal agreement between these predictions. The theoretical branching fractions are in the $\mathcal{O}(10^{-8} - 10^{-13})$ range \cite{gross, enter}, while the current experimental bounds are larger in the range of $\mathcal{O}(10^{-4} - 10^{-6})$ \cite{siru2,gaad, enter}. \\
\hspace*{1cm} In this work, by using the rare decay channels of  $W^{\pm} \rightarrow \pi^{\pm}\gamma$, $W^{\pm} \rightarrow K^{\pm}\gamma$, $W^{\pm} \rightarrow \rho^{\pm}\gamma$, we evaluate the contribution of new physics on the hadronic production at muon colliders in the RS model. This paper is arranged as follows. The theoretical framework is introduced in Section II. The contribution of new physics on the exclusive W boson hadronic decays in the final state at muon colliders are calculated in detail in Section III. Finally, we summarize our results and make conclusions in Section IV.
  \section{Theoretical framework of the Higgs - radion mixing in Randall-Sundrum model}
\hspace*{1cm}The RS model is based on a 5D spacetime with two three-branes: the UV brane and the IR brane. All the SM and dark matter (DM) fields, excepted for gravity, are assumed at IR brane. Gravity lives on the second three-branes. The RS model consider a non-factorizable 5-dimensional metric in the form \cite{dav} 
\begin{equation}
d_{s}^{2} = e^{-2 \sigma} \eta_{\mu \nu} d x^{\mu} d x^{\nu} - r_{c}^{2} d y^{2},
\end{equation}
where $\sigma = k r_{c}|y|$, $k$ is the curvature along the 5th-dimension, $r_{c}$ is the length-scale which is related to the size of the extra-dimension. For $\sigma=k r_c \sim 10$ the RS scenario can address the hierarchy problem. The action in $5 \mathrm{D}$ is
\begin{equation}
S=S_{\text {gravity }}+S_{\mathrm{IR}}+S_{\mathrm{UV}}.
\end{equation}
where 
\begin{equation}
S_{gravity} = \dfrac{16\pi}{M_{5}^{3}}\int d^{4}x\int^{\pi}_{0}r_{c}\sqrt{G^{(5)}} [R^{(5)}-2\Lambda_{5}] dy,
\end{equation}
$S_{\mathrm{IR}}, S_{\mathrm{UV}}$ are the brane actions for the two 3-branes as follows
\begin{equation}
S_{\mathrm{IR}} = \int d^{4}x\sqrt{-g}\left(-f^{4}_{IR} + \mathcal{L}_{SM} + \mathcal{L}_{DM}\right),
\end{equation}
\begin{equation}
S_{\mathrm{UV}} = \int d^{4}x\sqrt{-g}\left(-f^{4}_{UV} + ...\right).
\end{equation}
$M_{5}$ is the fundamental gravitational scale, $G^{(5)}_{MN}$ is the 5-dimensional metric, $R^{(5)}$ is the Ricci scalar, $\Lambda_{5}$ is the 5-dimensional cosmological constant, $f_{IR}, f_{UV}$ are the brane tensions for the two 3-branes and $f^{4}_{IR} = -f^{4}_{UV} = \sqrt{-24M^{3}_{5}\Lambda_{5}}$. On the IR brane, the effective gravitational coupling is $\Lambda = \overline{M}_{P} e^{-k \pi r_{c}}$. %For $\sigma = kr_{c} \sim 10$ the RS can address the hierarchy problem. 
The 4-dimensional component of the metric at first order is expanded
\begin{equation}
G^{(5)}_{MN} = e^{-2\sigma} (\eta_{\mu\nu} + \kappa_{5}h_{\mu\nu}),
\end{equation}
where $\kappa_{5} = 2M_{5}^{-2/3}$, the 5-dimensional field $h_{\mu\nu}$ can be written $h_{\mu\nu} (x, y) = \sum h_{\mu\nu}^{n}(x)\dfrac{\chi^{n}(y)}{\sqrt{r_{c}}}$. \\
\hspace*{1cm} The energy-momentum tensor is given by
\begin{equation}
T_{\mu \nu} = T_{\mu \nu}^{SM} + T_{\mu \nu}^{DM},
\end{equation}
where
\begin{equation}
\begin{aligned}
T_{\mu \nu}^{SM}= & {\left[\frac{i}{4} \bar{\psi}\left(\gamma_\mu D_\nu+\gamma_\nu D_\mu\right) \psi-\frac{i}{4}\left(\gamma_\mu D_\nu \bar{\psi} \gamma_\mu+D_\mu \bar{\psi} \gamma_\nu\right) \psi-\eta_{\mu \nu}\left(\bar{\psi} \gamma^\mu D_\mu \psi-m_\psi \bar{\psi} \psi\right)+\right.} \\
& \left.+\frac{i}{2} \eta_{\mu \nu} \partial^\rho \bar{\psi} \gamma_\rho \psi\right]+\left[\frac{1}{4} \eta_{\mu \nu} F^{\lambda \rho} F_{\lambda \rho}-F_{\mu \lambda} F_\nu^\lambda\right]+\left[\eta_{\mu \nu} D^\rho H^{\dagger} D_\rho H+\eta_{\mu \nu} V(H)+\right. \\
& \left.+D_\mu H^{\dagger} D_\nu H+D_\nu H^{\dagger} D_\mu H\right],
\end{aligned}
\end{equation}
and
\begin{equation}
T_{\mu \nu}^{DM}=\left(\partial_\mu S\right)\left(\partial_\nu S\right)-\frac{1}{2} \eta_{\mu \nu}\left(\partial^\rho S\right)\left(\partial_\rho S\right)+\frac{1}{2} \eta_{\mu \nu} m_S^2 S^2,
\end{equation}
\begin{equation}
T^{\mu}_{\mu}=\Sigma_{f} m_{f} \overline{f}f - 2m^{2}_{W}W^{+}_{\mu}W^{-\mu}-m^{2}_{Z}Z_{\mu}Z^{\mu} + (2m^{2}_{h_{0}}h_{0}^{2} -  \partial_{\mu}h_{0}\partial^{\mu}h_{0}) + ...
\end{equation}
The states that diagonalize the kinetic energy and have canonical normalization $h$ and $\phi$ are given by \cite{dominici}
\begin{equation} 
\left(\begin{array}{c} {h_{0} } \\ {\phi _{0} } \end{array}\right)=\left(\begin{array}
{cc} {1} & {6\xi \gamma /Z} \\ {0} & {-1/Z} \end{array}\right)\left(\begin{array}{cc}
 {\cos \theta } & {\sin \theta } \\ {-\sin \theta } & {\cos \theta } \end{array}\right)
 \left(\begin{array}{c} {h} \\ {\phi } \end{array}\right)=\left(\begin{array}{cc}
  {d} & {c} \\ {b} & {a} \end{array}\right)\left(\begin{array}{c} {h} \\ {\phi } \end{array}\right), \label{pt1}
\end{equation}
where
$Z^{2} = 1 + 6\gamma ^{2} \xi \left(1 -\, \, 6\xi \right) = \beta - 36\xi ^{2}\gamma ^{2}$ is the coefficient of the radion kinetic term after undoing the kinetic mixing, $\gamma = \upsilon /\Lambda _{\phi }, \upsilon = 246$ GeV, $a = -\dfrac{cos\theta}{Z}, b = \dfrac{sin\theta}{Z}, c = sin\theta + \dfrac{6\xi\gamma}{Z}cos\theta, d = cos\theta - \dfrac{6\xi\gamma}{Z}sin\theta$. The parameters a, b, c, d define the mixing between the $\xi = 0$ states and the $\xi \neq 0$ mass eigenstates \cite{boo}. \\
\hspace*{1cm}The mixing angle $\theta $ is
\begin{equation}
\tan 2{\theta } = 12{\gamma \xi Z}\frac{m_{h_{0}}^{2}}{m_{h_{0}}^{2} \left( Z^{2} - 36\xi^{2} \gamma ^{2} \right) - m_{\phi _{0}}^{2}},
\end{equation}
where $m_{h_{0}}$ and $m_{\phi _{0}}$ are the Higgs and radion masses before mixing. The new physical fields h and $\phi $ in (\ref{pt1}) are Higgs-dominated state and radion. The square masses are given by
\begin{equation} 
m_{h,\phi }^{2} =\frac{1}{2Z^{2} } \left[m_{\phi _{0} }^{2} +\beta m_{h_{0} }^{2} \pm \sqrt{(m_{\phi _{0} }^{2} +\beta m_{h_{0} }^{2} )^{2} -4Z^{2} m_{\phi _{0} }^{2} m_{h_{0} }^{2} } \right].
\end{equation}
\hspace*{1cm}There are four independent parameters $\Lambda _{\phi } ,\, \, m_{h} ,\, \, m_{\phi } ,\, \, \xi$ that must be specified to fix the state mixing parameters. We consider the case of  $\Lambda _{\phi } = 5$ TeV and $\frac{m_{0} }{M_{P} } = 0.1$ which makes the radion stabilization model most natural \cite{dav}.\\ 
\hspace*{1cm} A compelling question arises regarding whether a intuitive and tractable alternative exists for modeling unparticles. A natural theoretical candidate for this purpose is found in models based on warped extra dimensions. This idea is mentioned in RS model \cite{fried, iltan}. The effective Lagrangian can be written by 
\begin{equation}
\mathcal{L}_{eff} = \mathcal{L}_{SM} + \mathcal{L}_{RS} (h, \phi; \xi, \Lambda_{\phi}) + \frac{C_{UW}}{\Lambda_{U}^{d_{U}}} \mathcal{O}_{U}\mathcal{O}_{W} + \frac{C_{U \mu}}{\Lambda_{U}^{d_{U} - 1}} \mathcal{O}_{U}\overline{\mu}\mu + ...
\end{equation}
We use couplings of the SM particles with the Higgs, radion and unparticle listed in detail in Ref.\cite{ahm, cheung, chenhe, georgi2} for evaluating in the next section.

\section{The contribution of new physics on the exclusive W boson hadronic decays in the final state at muon colliders}
\hspace*{1cm} Influence of unparticle and polarization on properties of high energy colliders have been intensively studied in Refs.\cite{pra,alan,maj, kuma,sahi,kiku,chen,kha, fried, iltan, giang,soa1}. The production of dark matter and unparticles associated with Z boson at 8 TeV via pp collisions has been analyzed in Ref.\cite{khac}. Recently, the scalar unparticle signals at LHC 14 TeV are investigated in Ref.\cite{alie}. Phenomenology of heavy neutral gauge boson at muon colliders are presented in detail in \cite{ZLu} which indicate that it can provide an unprecedented potential in probing new physics beyond the SM. \\
\hspace*{1cm} Study for the exclusive boson hadronic decay with the ATLAS detector at 13 TeV is considered in detail in Ref.\cite{gaad}. Recently, mass biases in reconstructing exclusive radiative hadronic decays of W boson at the LHC are given in Ref.\cite{ejo}. While rare and exclusive few-body decays of the Higgs, Z and W bosons and the top quark are studied in Ref.\cite{enter}, the W boson's exotic decay channel is concerned in Ref.\cite{fei}. However, the contribution of unparticle and anomalous couplings at muon colliders through hadronic decays of W boson has not yet been invested.
In this section, by using the rare decay channels of $W^{\pm} \rightarrow \pi^{\pm}\gamma$, $W^{\pm} \rightarrow K^{\pm}\gamma$, $W^{\pm} \rightarrow \rho^{\pm}\gamma$, we investigate the influence of new physics on the hadronic production at muon colliders in the RS model.\\	

\begin{figure} [!htbp]
\begin{center}
\includegraphics[width= 15 cm,height= 4.5 cm]{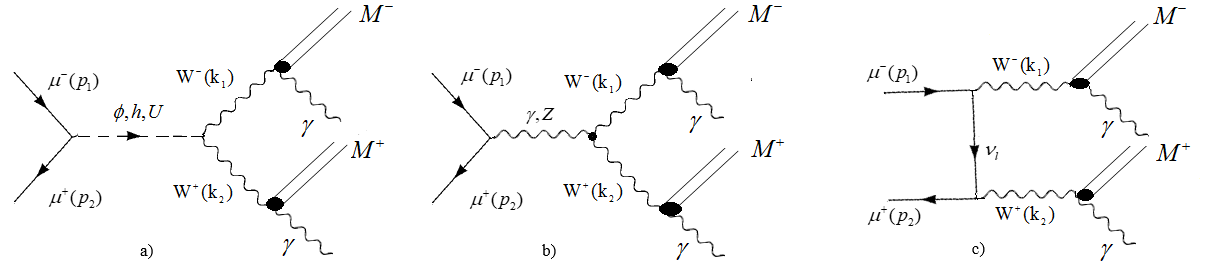}
\caption{\label{Fig.1} Feynman diagrams for $\mu^{+}\mu^{-} \rightarrow  W^{+}W^{-} \rightarrow \pi^{-}\pi^{+}\gamma\gamma/K^{-}K^{+}\gamma\gamma/\rho^{-}\rho^{+}\gamma\gamma$ collisions. $M^{\pm}$ stands for the $\pi^{\pm}, K^{\pm}, \rho^{\pm}$ .}
\end{center}
\end{figure}

\begin{figure}[!htb] 
\begin{center}
    \begin{tabular}{ccc}
        \includegraphics[width=5.5cm, height= 3.5cm]{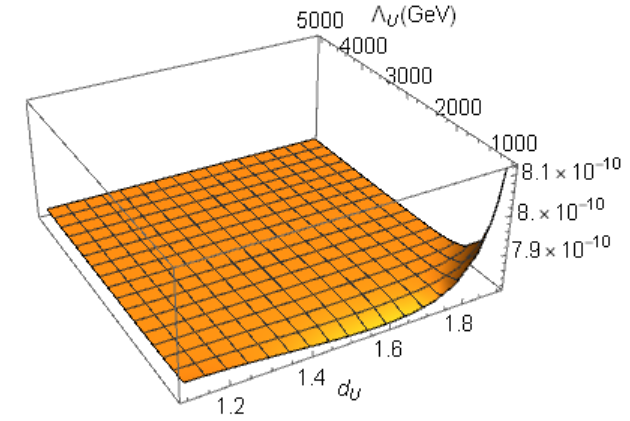} &
        \includegraphics[width=5.5cm, height= 3.5cm]{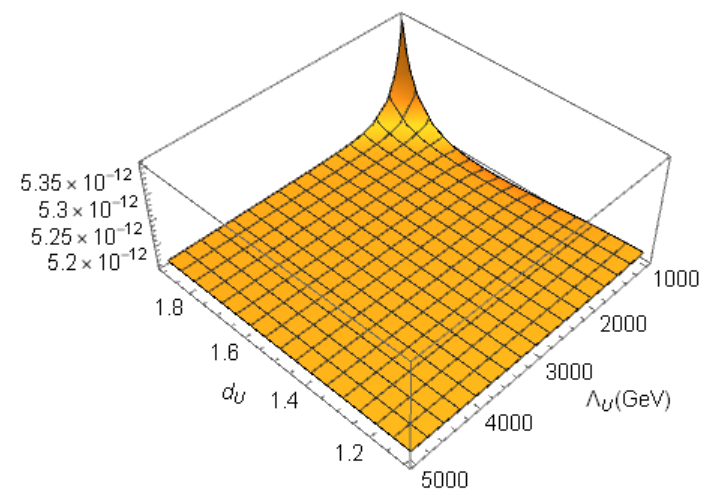} &        
         \includegraphics[width=5.5cm, height= 3.5cm]{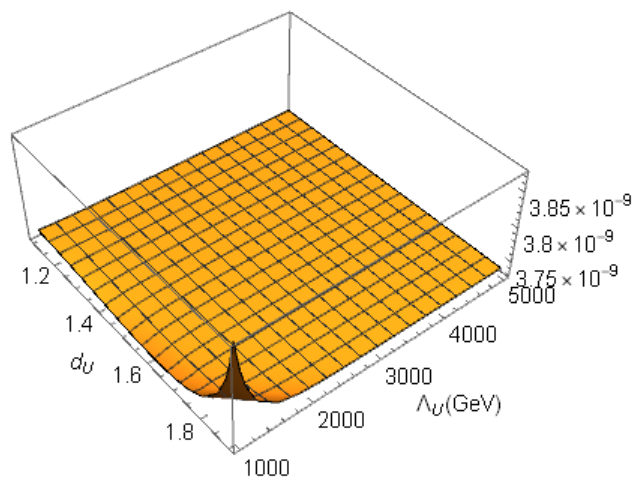}\\ 
         \small(a)& \small(b) & \small(c)
    \end{tabular}
    \caption{\label{Fig.2} The total cross-section depends on the ($\Lambda_{U}, d_{U}$) in the reactions of (a) $\mu^{+}\mu^{-} \rightarrow W^{+}W^{-} \rightarrow \pi^{+}\pi^{-}\gamma\gamma$, (b) $\mu^{+}\mu^{-} \rightarrow W^{+}W^{-} \rightarrow K^{+}K^{-}\gamma\gamma$ and (c) $\mu^{+}\mu^{-} \rightarrow W^{+}W^{-} \rightarrow \rho^{+}\rho^{-}\gamma\gamma$, respectively. The parameters are chosen as $\sqrt{s} = 10$ TeV,  $(P_{\mu^{-}}, P_{\mu^{+}} )= (0.8,-0.8)$, $(\Delta k_{\gamma},\lambda_{\gamma}) = (0.190, 0.061)$, $(\Delta k_{Z},\lambda_{Z}) = (-0.061,-0.062)$.} 
    \end{center}
\end{figure}

\begin{figure}[!htb] 
\begin{center}
    \begin{tabular}{ccc}
        \includegraphics[width=5.5cm, height= 3.5cm]{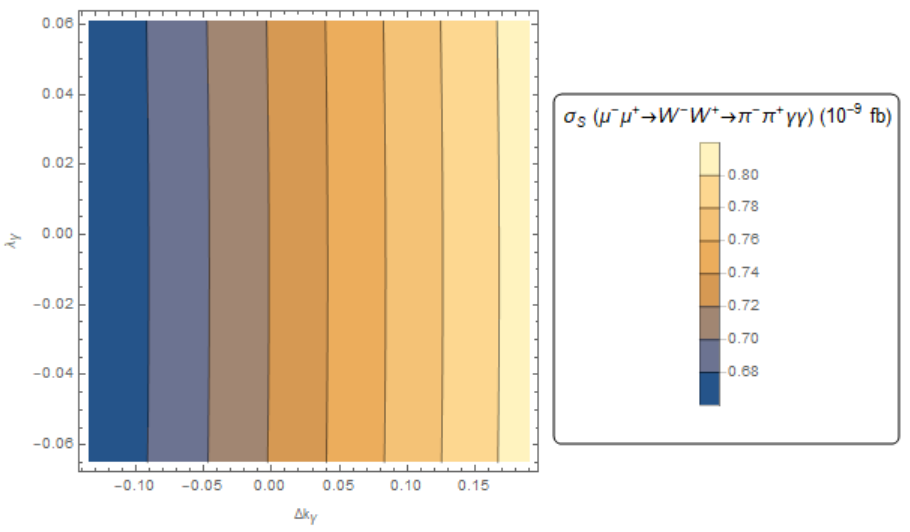} &
        \includegraphics[width=5.5cm, height= 3.5cm]{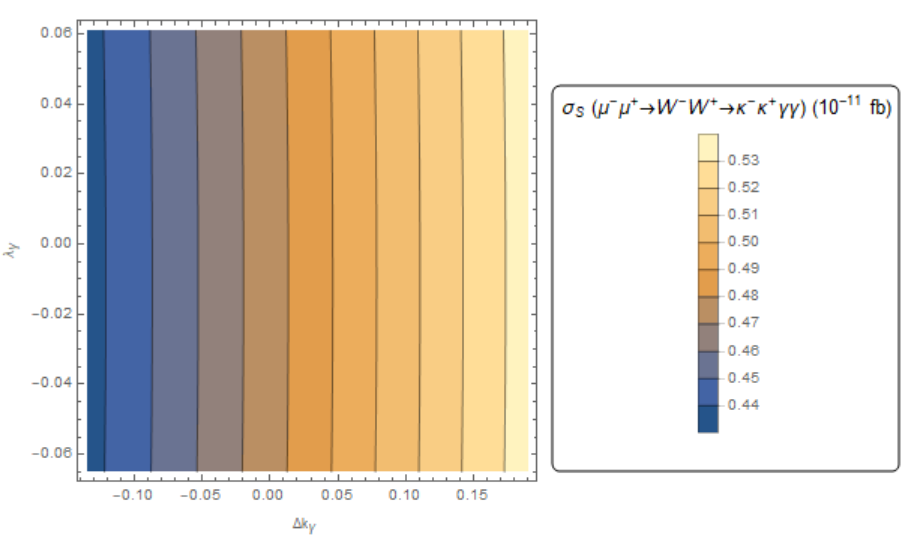} & 
         \includegraphics[width=5.5cm, height= 3.5cm]{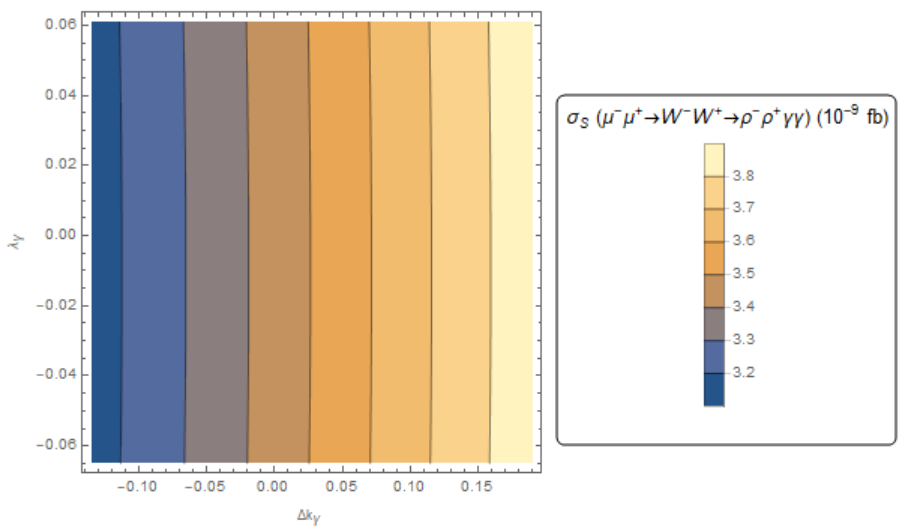}\\
        \small(a)& \small(b) & \small(c)
            \end{tabular}
    \caption{\label{Fig.3}  The total cross-section depends on the ($\Delta k_{\gamma}, \lambda_{\gamma}$). The benchmark signal point are taken to be $(\Lambda_{U}, d_{U})$ $= (1 \text{TeV}, 1.9)$. The other parameters are chosen as in Fig.\ref{Fig.2}.} 
    \end{center}
\end{figure}

\begin{figure}[!htb] 
\begin{center}
    \begin{tabular}{ccc}
        \includegraphics[width=5.5cm, height= 3.5cm]{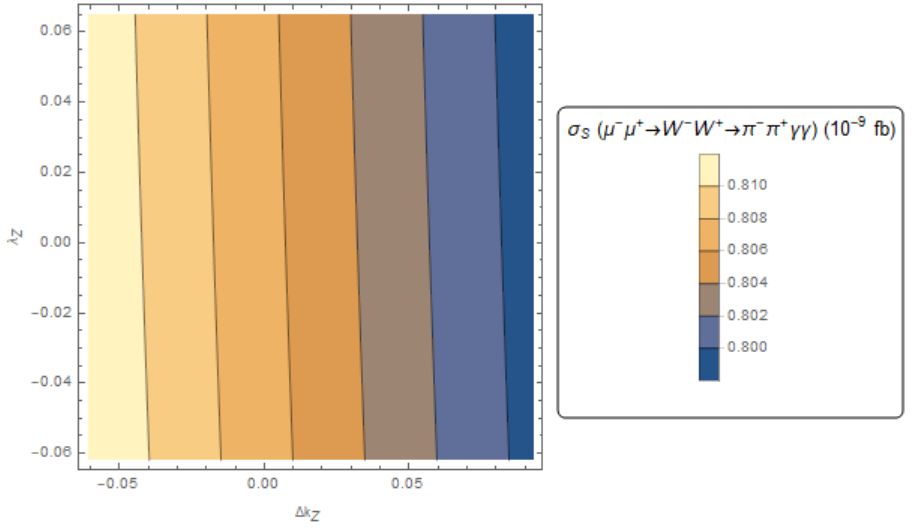} &
        \includegraphics[width=5.5cm, height= 3.5cm]{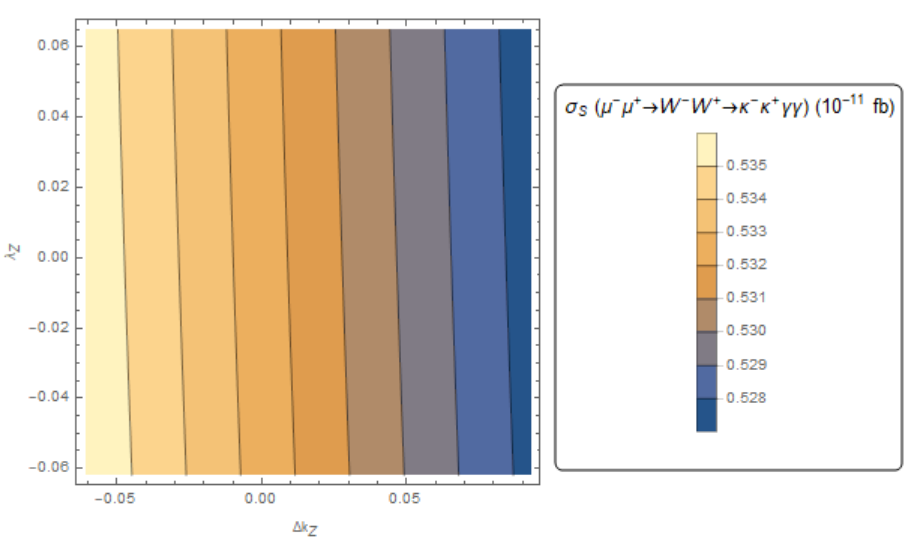} &
                \includegraphics[width=5.5cm, height= 3.5cm]{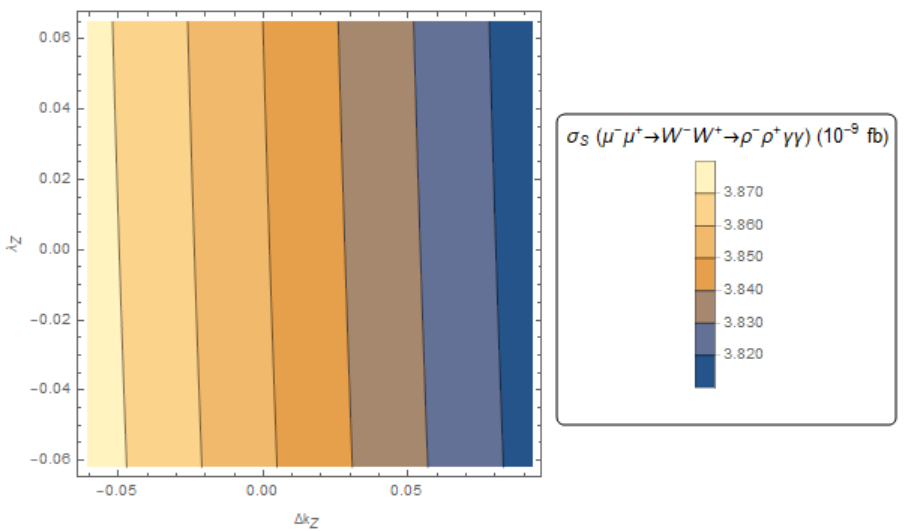}\\
       \small(a)& \small(b) & \small(c)
    \end{tabular}
    \caption{\label{Fig.4}  The total cross-section depends on the ($\Delta k_{Z}, \lambda_{Z}$). The benchmark signal point are taken to be $(\Lambda_{U}, d_{U})$ $= (1 \text{TeV}, 1.9)$. The parameters are chosen as in Fig.\ref{Fig.2}.} 
    \end{center}
\end{figure}

\begin{figure}[!htb] 
\begin{center}
    \begin{tabular}{ccc}
        \includegraphics[width=5.5cm, height= 3.5cm]{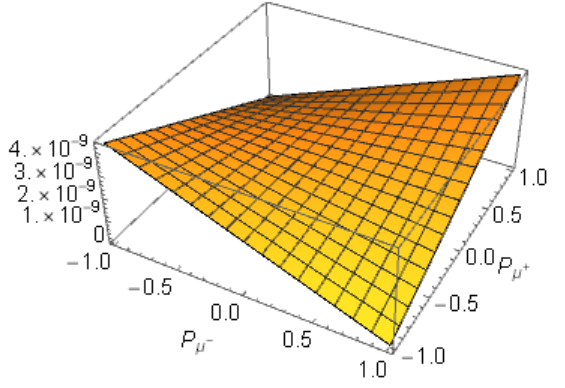} &
        \includegraphics[width=5.5cm, height= 3.5cm]{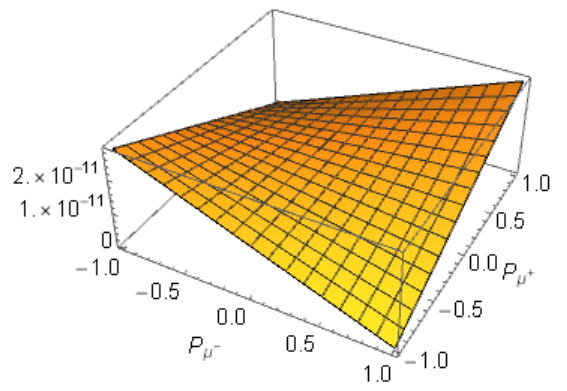} &         
         \includegraphics[width=5.5cm, height= 3.5cm]{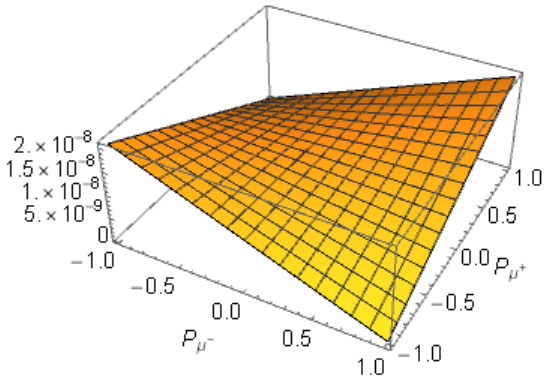} \\
                 \small(a)& \small(b) & \small(c)
           \end{tabular}
    \caption{\label{Fig.5} The total cross-section as a function of the polarization coefficients ($ P_{\mu^{-}}, P_{\mu^{+}}$). The parameters are chosen as $\sqrt{s} = 10$ TeV, $(\Lambda_{U}, d_{U})$ $= (1 \text{TeV}, 1.9)$, $(\Delta k_{\gamma},\lambda_{\gamma}) = (0.190, 0.061)$, $(\Delta k_{Z},\lambda_{Z}) = (-0.061,-0.062)$.}
    \end{center}
\end{figure}

\begin{figure}[!htb] \begin{center}
    \begin{tabular}{ccc}
        \includegraphics[width=4.5cm, height= 3.5cm]{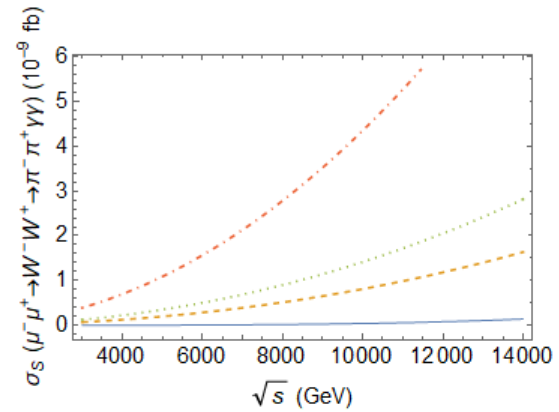} &
        \includegraphics[width=4.5cm, height= 3.5cm]{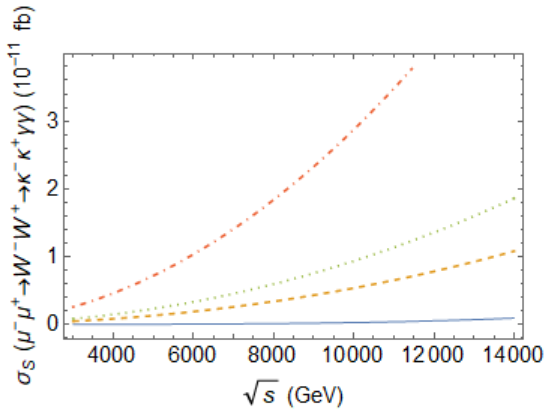} &
         \includegraphics[width=6.5cm, height= 3.5cm]{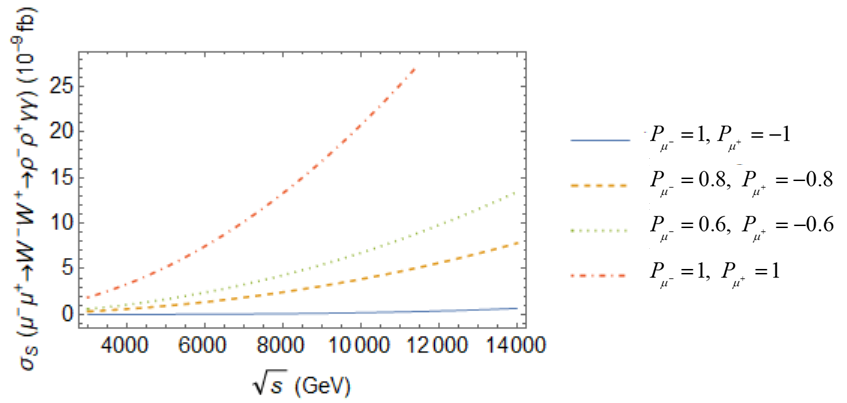}\\
         \small(a)& \small(b) & \small(c)
    \end{tabular}
    \caption{\label{Fig.6}  The total cross-section depends on the collision energy. The parameters are chosen as in Fig.\ref{Fig.5}.}
    \end{center}
\end{figure}

\hspace*{1cm}Now we consider the collision process $\mu^{+}\mu^{-} \rightarrow W^{+}W^{-}$,  
\begin{equation} \label{pt4}
\mu^{-}(p_{1}) + \mu^{+}(p_{2}) \    \rightarrow         W^{-} (k_{1}) + W^{+} (k_{2}).
\end{equation}
\hspace*{1cm}The transition amplitude representing s-channel is given by
\begin{equation}
M_{s} = M_{\gamma} + M_{Z} + M_{\phi} + M_{h} + M_{U},
\end{equation}
where
\begin{equation}
M_{\gamma} =  i \dfrac{e}{q^{2}_{s}}\overline{v}(p_{2})\gamma^{\sigma} u(p_{1}) \eta_{\sigma \beta} \varepsilon^{*}_{\mu} (k_{1}) \Gamma_{WW\gamma}^{\beta\mu\nu} \varepsilon^{*}_{\nu} (k_{2}),
\end{equation}
\begin{equation}
M_{Z} = -i \dfrac{g}{2 \cos \theta_{W} (q^{2}_{s} - m^{2}_{Z})}\overline{v}(p_{2}) \gamma^{\sigma} \left(-\frac{1}{2} + 2 \sin^{2}\theta_{W} - \frac{1}{2} \gamma_{5}\right) u(p_{1}) \left(\eta_{\sigma\beta} - \dfrac{q_{s\sigma}q_{s\beta}}{m^{2}_{Z}} \right) \varepsilon^{*}_{\mu} (k_{1})\Gamma_{WWZ}^{\beta\mu\nu} \varepsilon^{*}_{\nu} (k_{2})
\end{equation}
\begin{equation}
M_{\phi} = -i \dfrac{\overline{g}_{\mu\mu\phi}\overline{g}_{W\phi}}{q^{2}_{s} - m^{2}_{\phi}}\overline{v}(p_{2})u(p_{1}) \varepsilon^{*}_{\mu} (k_{1}) \left[\eta^{\mu\nu} - 2g^{W}_{\phi}\left(\left(k_{1}k_{2}\right)\eta^{\mu\nu} - k_{1}^{\nu}k_{2}^{\mu}\right)\right]\varepsilon^{*}_{\nu} (k_{2}),
\end{equation}
\begin{equation}
M_{h} = - i\dfrac{\overline{g}_{\mu\mu h}\overline{g}_{Wh}}{q^{2}_{s} - m^{2}_{h}}\overline{v}(p_{2})u(p_{1}) \varepsilon^{*}_{\mu} (k_{1}) \left[\eta^{\mu\nu} - 2g^{W}_{h}\left(\left(k_{1}k_{2}\right)\eta^{\mu\nu} - k_{1}^{\nu}k_{2}^{\mu}\right)\right]\varepsilon^{*}_{\nu} (k_{2}),
\end{equation}
\begin{equation}
M_{U} = i\overline{g}_{\mu\mu U}\overline{g}_{WWU} \dfrac{A_{d_{U}}}{2sin(d_{U}\pi)} |q^{2}_{s}|^{d_{U} - 2} e^{-i\pi (d_{U} - 2)} \overline{v}(p_{2})u(p_{1})\varepsilon^{*}_{\mu} (k_{1}) \left[\left(k_{1}k_{2}\right)\eta^{\mu\nu} - k_{1}^{\nu}k_{2}^{\mu}\right]\varepsilon^{*}_{\nu} (k_{2}).
\end{equation}
Here, $q_{s} = p_{1} + p_{2} = k_{1} + k_{2}$, $s = (p_{1} + p_{2})^{2}$ is the square of the collision energy, $M_{U}$ is the contribution by the scalar unparticle, which is important for the described process. $\overline{g}_{\mu\mu\phi}, \overline{g}_{\mu\mu h}, \overline{g}_{W\phi}, \overline{g}_{Wh}$ are the couplings of Higgs/radion shown in Ref.\cite{ahm}. $\overline{g}_{\mu\mu U}, \overline{g}_{WWU}, A_{d_{U}}$ are given by Ref.\cite{cheung, georgi2}. Anomalous couplings are shown in Ref.\cite{bhatia} as follows\\
\begin{equation}
i\Gamma^{WW\gamma}_{\mu\nu\lambda}(p_{1},p_{2},p_{3}) = ie\left[T_{\mu\nu\lambda}^{(0)}(p_{1},p_{2},p_{3}) + \Delta k_{\gamma}T_{\mu\nu\lambda}^{(1)}(p_{1},p_{2},p_{3}) + \dfrac{\lambda_{\gamma}}{M_{W}^{2}}T_{\mu\nu\lambda}^{(2)}(p_{1},p_{2},p_{3})\right],
\end{equation}
\begin{equation}
i\Gamma^{WWZ}_{\mu\nu\lambda}(p_{1},p_{2},p_{3}) = ie\left[T_{\mu\nu\lambda}^{(0)}(p_{1},p_{2},p_{3}) + \Delta k_{Z}T_{\mu\nu\lambda}^{(1)}(p_{1},p_{2},p_{3}) + \dfrac{\lambda_{Z}}{M_{W}^{2}}T_{\mu\nu\lambda}^{(2)}(p_{1},p_{2},p_{3})\right],
\end{equation}
where the $T_{\mu\nu\lambda}$ tensors are, respectively,
\begin{subequations}
\begin{align}
&T_{\mu\nu\lambda}^{(0)} = \eta_{\mu\nu}(p_{1}-p_{2})_{\lambda} + \eta_{\nu\lambda}(p_{2}-p_{3})_{\mu} + \eta_{\lambda\mu}(p_{3}-p_{1})_{\nu},\\
&T_{\mu\nu\lambda}^{(1)} = \eta_{\lambda\mu}p_{3\nu} - \eta_{\nu\lambda}p_{3\mu},\\
&T_{\mu\nu\lambda}^{(2)} = p_{1\lambda}p_{2\mu}p_{3\nu} - p_{1\nu}p_{2\lambda}p_{3\mu} - \eta_{\mu\nu}(p_{2}p_{3} p_{1\lambda} - p_{3}p_{1}p_{2\lambda}) - \eta_{\nu\lambda}(p_{3}p_{1} p_{2\mu} - p_{1}p_{2}p_{3\mu}) \eta_{\mu\lambda}(p_{1}p_{2} p_{3\nu} - p_{2}p_{3}p_{1\nu}).  
\end{align}
\end{subequations}
The parameters $\Delta \kappa_{V}$ and $\lambda_{V} (V = \gamma, Z)$ serve to quantify the strength of the beyond SM physics. Experiment results from W boson characterization constrain $\Delta \kappa_{V}$, indicating that these corrections cannot exceed a few percent. The parameter $\lambda_{V}$ can be identified as follows: $\lambda_{V} \rightarrow \lambda_{V}/(1 + s/\Lambda^{2})^{2}$ \cite{navas}.\\
\hspace*{1cm} The transition amplitude representing t-channel can be written as
\begin{equation}
M_{t} = -\frac{g^{2}}{2} \varepsilon^{*}_{\nu} (k_{2}) \overline{v}(p_{2}) \gamma^{\mu} P_{L} \frac{\widehat{q}_{t}}{q^{2}_{t} - m^{2}_{\nu}} \varepsilon^{*}_{\mu} (k_{1}) \gamma^{\nu} P_{L} u(p_{1}).
\end{equation}
\hspace*{1cm}
 The total cross-section for the whole process can be calculated as follows
\begin{align}
&\sigma_{\pi^{-}\pi^{+}\gamma\gamma} = \sigma (\mu^{-} \mu^{+} \rightarrow  W^{+}W^{-}) \times   Br(W^{-}\rightarrow \pi^{-}\gamma) Br(W^{+} \rightarrow \pi^{+}\gamma),\\
&\sigma_{K^{-}K^{+}\gamma\gamma} = \sigma (\mu^{-} \mu^{+} \rightarrow  W^{+}W^{-}) \times   Br(W^{-}\rightarrow K^{-}\gamma) Br(W^{+} \rightarrow K^{+}\gamma),\\
&\sigma_{\rho^{-}\rho^{+}\gamma\gamma} = \sigma (\mu^{-} \mu^{+} \rightarrow  W^{+}W^{-}) \times   Br(W^{-}\rightarrow \rho^{-}\gamma) Br(W^{+} \rightarrow \rho^{+}\gamma).
\end{align}
From the expressions of the differential cross-section \cite{pes}
\begin{equation}
\frac{d\sigma (\mu^{-} \mu^{+} \rightarrow  W^{+}W^{-})}{dcos\psi} = \frac{1}{32 \pi s} \frac{|\overrightarrow{k}_{1}|}{|\overrightarrow{p}_{1}|} |M_{fi}|^{2},
\end{equation}
where $\psi = (\overrightarrow{p}_{1}, \overrightarrow{k}_{1})$ is the scattering angle. The model parameters are chosen as $\lambda_{\mu \mu} = \lambda_{WW} = \lambda_{0} = 1$, $\xi = 1/6$ \cite{ahm}, $\Lambda_{\phi} = 5$ TeV \cite{dav}, $ m_{h}$ = 125 GeV, $m_{\phi}$ = 125 GeV \cite{giang2023}. The integrated luminosity scaling of a high energy muon collider reaches 10 $\text{ab}^{-1}$ at a center-of-mass energy of $\sqrt{s} = 10$ TeV \cite{wulzer, liu}. The bounds on the anomalous $W^{-}W^{+}\gamma$ and $W^{-}W^{+}Z$ couplings are provided by the LEP, Tevatron and LHC experiments. The ATLAS collaboration has updated the best available constraints on anomalous couplings $\Delta k_{\gamma}$, $\lambda_{\gamma}$, $\Delta k_{Z}$, $\lambda_{Z}$ obtained as follows: $\Delta k_{\gamma} \in [-0.135, 0.190]$, $\lambda_{\gamma} \in [-0.065, 0.061]$, $\Delta k_{Z} \in [-0.061, 0.093]$, $\lambda_{Z}\in [-0.062, 0.065]$ \cite{cakir}.
 From the formulas (28 - 30) above and branching ratio of W boson in the soft-collinear effective theory (SCET) and  the light-cone distribution amplitude (LCDA) framework shown in Ref.\cite{enter}, we give estimates in detail for the cross-sections from Fig.\ref{Fig.2} to Fig.\ref{Fig.6} as follows:\\
\hspace*{0.5cm} Initially, we evaluate the benchmark parameters $(\Lambda_{U}, d_{U})$ in Fig.\ref{Fig.2}. The parameters are taken to be $P_{\mu^{-}} = 0.8, P_{\mu^{+}} = -0.8$, $\sqrt{s} = 10$ TeV, $\Delta k_{\gamma} = 0.190$, $\lambda_{\gamma} = 0.061$, $\Delta k_{Z} = -0.061$, $\lambda_{Z} = -0.062$. From the figure, we can see that the cross-sections reach their maximum values at the benchmark signal point $(\Lambda_{U}, d_{U})$ $= (1 \text{TeV}, 1.9)$.\\ 
 \hspace*{0.5cm} Furthermore, with the benchmark signal point $(\Lambda_{U}, d_{U}) = (1 \text{TeV}, 1.9) $, the total cross-section depends on ($\Delta k_{\gamma}$, $\lambda_{\gamma}$) shown in the Fig. \ref{Fig.3}. The other parameters are chosen as in Fig.\ref{Fig.2}. The figure shows that cross-sections are the largest in the yellow region, the typical values are given by (a) $0.8 \times 10^{-9}$ fb for the decay $W^{\pm} \rightarrow \pi^{\pm}\gamma$, (b) $0.53 \times 10^{-11}$ fb for the decay $W^{\pm} \rightarrow K^{\pm}\gamma$ and (c) $3.8 \times 10^{-9}$ fb for the decay $W^{\pm} \rightarrow \rho^{\pm}\gamma$, respectively. \\

 \hspace*{0.5cm} Similar to the Fig.\ref{Fig.3}, the total cross-sections depend on ($\Delta k_{Z}$, $\lambda_{Z}$) are shown in the Fig.\ref{Fig.4}. The typical values of the cross -section are given by (a) $0.810 \times 10^{-9}$ fb, (b) $0.535 \times 10^{-11}$ fb and (c) $3.870 \times 10^{-9}$ fb, respectively.\\

  \hspace*{0.5cm} In Fig.\ref{Fig.5}, the cross-sections are plotted as the function of polarization coefficients ($P_{\mu^{-}}, P_{\mu^{+}}$). The parameters are chosen as above, i.e  $d_{U} = 1.9$, $\Lambda_{U} = 1$ TeV, $\sqrt{s} = 10$ TeV, $\Delta k_{\gamma} = 0.190$, $\lambda_{\gamma} = 0.061$, $\Delta k_{Z} = -0.061$, $\lambda_{Z} = -0.062$. The figures indicate that the cross-sections achieve the maximum values when $P_{\mu^{-}} = P_{\mu^{+}} = 1$ or $-1$ and the minimum values when $P_{\mu^{-}} = 1, P_{\mu^{+}} = -1$ or $P_{\mu^{-}} = -1, P_{\mu^{+}} = 1$, respectively. \\
  
\hspace*{0.5cm} Next, we evaluate the dependence of the cross-section on the energy, as shown in Fig.\ref{Fig.6}. The parameters are chosen as in Fig.\ref{Fig.5} and  the polarization coefficients $(P_{\mu^{-}}, P_{\mu^{+}})$ are similar to  Ref.\cite{ZLu}, i.e  $(P_{\mu^{-}}, P_{\mu^{+}})$ $= (1, -1), (0.8, -0.8), (0.6, -0.6), (1, 1)$, respectively. From the figure we can see that the cross-sections increase when the collision energy increases. In case of $\mu^{-}$ beam is left polarized, $\mu^{+}$ beam is right polarized and vice versa, the cross-sections change insignificantly. \\
 
 \hspace*{0.5cm} For more details, we calculate the statistical significance for the exclusive W boson hadronic decays at muon colliders which is defined by \cite{chakra, cxyue} $SS = S/\sqrt{S + B}$, where B, S are the number of events for the signal in SM framework and in RS framework including unparticle, anomalous couplings, respectively. The parameters are chosen as in Fig.\ref{Fig.5} and the polarization coefficients of $(P_{\mu^{-}}, P_{\mu^{+}} )= (1,1)$ are chosen. Using the integrated luminosity scaling of 10 $\text{ab}^{-1}$ (assuming a 5-year run) \cite{wulzer, liu}, the some typical values of the statistical significance with different values of the W branching ratio in SCET and LCDA framework in Ref.\cite{enter} are given in detail in Tables 1 and in Table 2, respectively. From Table 1 we can see that at the theoretical bound of $\mathcal{O}(10^{-9})$, the statistical significance remains small. While in Table 2, using the W branching ratio at the experimental bounds of $\mathcal{O} (10^{-6})$\cite{gaad} the statistical significance is  given by, $\text{SS} = (1.4444 ^{+0.6018}_{-0.3611})\sigma$, $(1.3239^{+0.4815}_{-0.3611})\sigma$ and  $(7.2218^{+2.7684}_{-2.0463})\sigma$ for the decays of $W^{\pm} \rightarrow \pi^{\pm}\gamma$, $W^{\pm} \rightarrow K^{\pm}\gamma$ and $W^{\pm} \rightarrow \rho^{\pm}\gamma$, respectively. This result shows that with the contribution the new physics from the RS model including unparticle and anomalous couplings the effect is enhanced and can be measured at the multi-TeV muon collider.

%%%%%%%%%%%%%%%%%%%%%%%%%%%%%%%%%%%%%%%%%%%%%%%%%%%%%%%

 \begin{table}[!htb]
 	  	  \caption{\label{tab1} Some typical values of the statistical significance for the exclusive W boson hadronic decays at high energy muon colliders with the integrated luminosity of 10 $\text{ab}^{-1}$ in case of using the W branching ratio in the SCET and LCDA framework in Ref.\cite{enter}, $\mathcal{O}(10^{-9})$. The parameters are chosen as in Fig.\ref{Fig.5} and the polarization coefficients of $(P_{\mu^{-}}, P_{\mu^{+}} )= (1,1)$ are chosen.}
 	  	  
 	  	  \vspace*{0.25cm}
 	  	\begin{center}
 	   	  \begin{tabular}{|c|c|c|c|}     	 
 	   	  \hline
 	   	  Process &  $\sigma_{B}$ (fb) & $\sigma_{S}$ (fb) & $SS(\sigma)$  \\
	\hline 
				$\mu^{+}\mu^{-} \rightarrow W^{+}W^{-} \rightarrow \pi^{-} \pi^{+} \gamma\gamma$ & (3.823 $\pm$ 0.153 )  $\times 10^{-9}$ & (4.354 $\pm 0.174$) $\times 10^{-9}$ & 0.0048 $\pm$ 0.00096  \\
		\hline
				$\mu^{+}\mu^{-} \rightarrow W^{+}W^{-} \rightarrow K^{-} K^{+} \gamma\gamma$ & (2.524 $\pm$ 0.114) $\times 10^{-11}$ & (2.874  $\pm 0.129 $) $\times 10^{-11}$ &  0.00039 $\pm$ 0.00008 \\
			\hline
					$\mu^{+}\mu^{-} \rightarrow W^{+}W^{-} \rightarrow \rho^{-} \rho^{+} \gamma\gamma$ & (18.252 $\pm$ 0.872 ) $\times 10^{-9}$ & (20.785 $\pm 0.993 $) $\times 10^{-9}$ & 0.0105 $\pm$ 0.00229  \\
			\hline
			 	 \end{tabular}
    \end{center}
   \end{table}
   \begin{table}[!htb]
 	  	  \caption{\label{tab2} Some typical values in the layers: SM, RS including unparticle and anomalous couplings in case of using the W branching ratio in the SCET and LCDA framework in Ref.\cite{enter}, $\mathcal{O}(10^{-9})$. The parameters are chosen as in Fig.\ref{Fig.5} and the polarization coefficients of $(P_{\mu^{-}}, P_{\mu^{+}} )= (1,1)$ are chosen.}
 	  	  
 	  	  \vspace*{0.25cm}
 	  	\begin{center}
 	   	  \begin{tabular}{|c|c|c|c|}     	 
 	   	  \hline
 	   	Final states &  $\sqrt{s}$ &  $\sigma_{B} (\sigma_{SM})$ (fb) & $\sigma_{S} (\sigma_{\text{RS + U + anomalous}})$ (fb)   \\
	\hline 
	$\pi^{-} \pi^{+} \gamma\gamma$ & 3 &  (3.450 $\pm$ 0.138 ) $\times 10^{-10}$ & (3.929 $\pm 0.157$) $ \times 10^{-10}$  \\
		%\hline
				&10 &  (3.823 $\pm$ 0.153 )  $\times 10^{-9}$ & (4.354 $\pm 0.174$) $\times 10^{-9}$  \\
		%	\hline
					&14 &  (7.526 $\pm$ 0.301 ) $\times 10^{-9}$ & (8.566 $\pm 0.343$) $\times 10^{-9}$   \\
			\hline
			$K^{-} K^{+} \gamma\gamma$ &  3 & (2.278 $\pm$ 0.103  ) $\times 10^{-12}$ & (2.594 $\pm 0.117$) $\times 10^{-12}$ \\
			& 10 & (2.524 $\pm$ 0.114) $\times 10^{-11}$ & (2.874  $\pm 0.129 $) $\times 10^{-11}$  \\
			& 14 & (4.968 $\pm$ 0.224 ) $\times 10^{-11}$ & (5.655 $\pm 0.255$) $\times 10^{-11}$  \\
			\hline
			$\rho^{-} \rho^{+} \gamma\gamma$ &  3 & (16.471 $\pm$ 0.787 ) $\times 10^{-10}$ & (18.757 $\pm 0.895 $) $\times 10^{-10}$  \\
			& 10 & (18.252 $\pm$ 0.872 ) $\times 10^{-9}$ & (20.785 $\pm 0.993 $) $\times 10^{-9}$ \\
			& 14 & (35.929 $\pm$ 1.716 ) $\times 10^{-9}$ & (40.894 $\pm 1.953$) $\times 10^{-9}$\\
			\hline			
			 	 \end{tabular}
    \end{center}
   \end{table}
 \begin{table}[!htb]
 	  	  \caption{\label{tab3} Some typical values of the statistical significance for the exclusive W boson hadronic decays at high energy  muon colliders with the integrated luminosity of 10 $\text{ab}^{-1}$ in case of using the W branching ratio at the experimental bounds of $\mathcal{O} (10^{-6})$ in Ref.\cite{gaad}. The parameters are chosen as in Fig.\ref{Fig.5} and the polarization coefficients of $(P_{\mu^{-}}, P_{\mu^{+}} )= (1,1)$ are chosen.}
 	  	  
 	  	  \vspace*{0.25cm}
 	  	\begin{center}
 	   	  \begin{tabular}{|c|c|c|c|}     	 
 	   	  \hline
 	   	  Process &  $\sigma_{B}$ (fb) & $\sigma_{S}$ (fb) & $SS(\sigma)$  \\
	\hline 
	& & & \\
					$\mu^{+}\mu^{-} \rightarrow W^{+}W^{-} \rightarrow \pi^{-} \pi^{+} \gamma\gamma$ & ($3.4407 ^{+0.5973}_{-0.2150}$ )  $\times 10^{-4}$ & (${3.9182}^{+ 0.6802}_{-0.2449}$) $\times 10^{-4}$
					& $1.4444 ^{+0.6018}_{-0.3611}$  \\
					& & &\\								 
		\hline
		& & &\\
				$\mu^{+}\mu^{-} \rightarrow W^{+}W^{-} \rightarrow K^{-} K^{+} \gamma\gamma$ & ($2.8911 ^{+0.3823}_{-0.2150}$) $\times 10^{-4}$ & ($3.2923 ^ {+0.4354} _{-0.2449}$) $\times 10^{-4}$ &  $1.3239^{+0.4815}_{-0.3611}$ \\
				& & &\\
			\hline
			& & &\\
					$\mu^{+}\mu^{-} \rightarrow W^{+}W^{-} \rightarrow \rho^{-} \rho^{+} \gamma\gamma$ & ($8.6016^{+1.2639}_{-0.6905} $) $\times 10^{-3}$ & ($9.7954 ^{+1.4394}_{-0.7864}$) $\times 10^{-3}$ & $7.2218^{+2.7684}_{-2.0463}$  \\
					& & &\\
			\hline
			 	 \end{tabular}
    \end{center}
   \end{table}
 \hspace*{0.5cm} Finally, to clarify the contribution of  the anomalous couplings $WW(\gamma,Z)$, we use a $\chi^{2}$ analysis with systematic errors to determine the sensitivities of the new contributions at the multi-TeV muon colliders. The $\chi^{2}$ function is defined as follows \cite{spor, koksal}
 \begin{equation}
 \chi^{2} = \left(\dfrac{\sigma_{SM} - \sigma_{NP}}{\sigma_{SM} \sqrt{(\delta_{st})^{2}+(\delta_{sys})^{2}}}\right)^{2},
\end{equation}     
where $\sigma_{SM}$ is the cross-section of SM backgrounds, $\sigma_{NP}$ is the cross-section containing contributions from presence of both new physics beyond the SM and SM backgrounds, defined by $\sigma_{NP} = \sigma_{SM} + \sigma_{INT} + \sigma_{BSM}$. Here, $\sigma_{INT}$ is the interference term between the SM and the new physics contribution and $\sigma_{BSM}$ is the contribution due to BSM physics, respectively. $\delta_{st} = \dfrac{1}{\sqrt{N_{SM}}} = \dfrac{1}{\sigma_{SM} \times \mathcal{L}}$ and $\delta_{sys}$ are the statistical error and the systematic error, respectively. $N_{SM} = \sigma_{SM} \times \mathcal{L}$ is the number of events of SM backgrounds, $\mathcal{L}$ is the integrated luminosity. The parameters are chosen as in Table.\ref{tab2}. The systematic error is chosen as $\delta_{sys} = 5\%$ \cite{spor}. Some typical values of $\chi^{2}$ for the contributions of the anomalous coupling $WW(\gamma,Z)$ are evaluated in case of using the W branching ratio of $\mathcal{O} (10^{-6})$ in Table.\ref{tab4}. The sensitivities to scalar propagators (unparticle, radion and Higgs) are found to be negligible. The best sensitivities obtained from the processes $\mu^{+}\mu^{-} \rightarrow  W^{+}W^{-} \rightarrow \pi^{-}\pi^{+}\gamma\gamma/K^{-}K^{+}\gamma\gamma/\rho^{-}\rho^{+}\gamma\gamma$ on the anomalous coupling $WW\gamma$  are much larger than that of the anomalous coupling $WW Z$ under the same conditions.

   \begin{table}[!htb]
 	  	  \caption{\label{tab4} Some typical values of $\chi^{2}$ for the contributions of photon, Z boson propagators in case of using the W branching ratio at the experimental upper bounds of $\mathcal{O} (10^{-6})$  in Ref.\cite{gaad}. The parameters are chosen as in Table.\ref{tab2}. The systematic error is chosen as $\delta_{sys} = 5\%$.}
 	  	   	  	  \vspace*{0.25cm}
 	  	\begin{center}
 	   	  \begin{tabular}{|c|c|c|c|c|}     	 
 	   	  \hline
 	   	 Final states & $\sqrt{s}$ (TeV) & $\mathcal{L}$ ($ab^{-1}$) & $\chi_{\gamma}^{2}$  & $\chi_{Z}^{2}$  \\
	\hline 
$\pi^{-} \pi^{+} \gamma\gamma$ &	3 & 1 & 2.3641 $\times 10^{-4}$ &5.9326 $\times 10^{-7}$ \\
					&10 & 10 &  2.5246 $\times 10^{-2}$ & 6.1945 $\times 10^{-5}$ \\		
						&14 & 20 &  9.3737 $\times 10^{-2}$ & 2.2975 $\times 10^{-4}$\\		
			\hline
			$K^{-} K^{+} \gamma\gamma$ &	3 & 1 &  1.8406 $\times 10^{-4}$ & 4.6189 $\times 10^{-7}$ \\
					&10 & 10 & 1.9729 $\times 10^{-2}$ & 4.8409 $\times 10^{-5}$ \\		
						&14 & 20 &  7.4022 $\times 10^{-2}$ & 1.8143 $\times 10^{-4}$\\		
			\hline
			$\rho^{-} \rho^{+} \gamma\gamma$  &	3 & 1 & 5.6155 $\times 10^{-3}$ & 1.4092 $\times 10^{-5}$ \\
					&10 & 10 & 4.3371 $\times 10^{-1}$ & 1.0642 $\times 10^{-3}$ \\		
						&14 & 20 &  9.1075 $\times 10^{-1}$ & 2.2322 $\times 10^{-3}$\\		
			\hline
			 	 \end{tabular}
    \end{center}
   \end{table}

\section{Conclusion}

\hspace*{1cm} In this paper, by using Feynman diagram techniques we have evaluated the influence of the scalar unparticle and polarization on the exclusive W boson hadronic decays of $W^{\pm} \rightarrow \pi^{\pm}\gamma$, $W^{\pm} \rightarrow K^{\pm}\gamma$ and  $W^{\pm} \rightarrow \rho^{\pm}\gamma$ at the high energy muon colliders in the RS model. The results show that with fixed collision  energies, the total cross-sections for hadronic productions in final state depend strongly on the parameters of the unparticle physics, muon beam polarizes and the anomalous couplings. With a center-of-mass energy of 10 TeV, the total cross-sections achieve the maximum value when the benchmark signal point as $(\Lambda_{U}, d_{U})$ $= (1 \text{TeV}, 1.9)$ and  the polarization coefficients as $(P_{\mu^{-}}, P_{\mu^{+}} )= (1,1)$.\\
\hspace*{1cm} For more details we have evaluated  the statistical significance for the considered process. The numerical values indicate that with the contribution of the new physics from the RS model including unparticle and anomalous couplings at the experimental bound of the W branching ratio of $\mathcal{O}(10^{-6})$, the effect is much enhanced and can be measured at the multi-TeV muon collider. The best sensitivities on the anomalous coupling $WW\gamma$ can reach at ($10^{-1}$ - $10^{-2}$) which are much larger than that on the anomalous coupling $WW Z$ under the same conditions.\\
\hspace*{1cm} Finally, we emphasize that in this work we have only considered on a theoretical basis, other problems concerning experiments of exclusive boson hadronic decays, the reader can see in detail in Refs.\cite{gaad,ejo}. It is worth noting that these rare processes provide a test bench for the quantum chromodynamics factorization formalism used to calculate cross-sections at high energy colliders, as well as a probe of W boson coupling to quarks and a new way to measure the W boson mass through fully reconstructed decay products. We will continue to investigate processes resulting in the decay channels of W boson to other meson final states in our subsequent studies.\\		
\\
{\bf Acknowledgements}: The work is supported in part by the National Foundation for Science and Technology Development (NAFOSTED) of Vietnam under Grant No. 103.01-2023.50.\\

\end{document}